\documentclass[journal,hideappendix]{main/vgtc}        % final (journal style) without appendices
\onlineid{1603}

\vgtccategory{Research}

\usepackage{enumitem}
\usepackage{quoting}
\quotingsetup{vskip=1pt}
\usepackage[final]{changes}
\usepackage{xcolor}
\usepackage[normalem]{ulem}
\usepackage{fancyvrb}

\setaddedmarkup{\textcolor{red}{#1}}

\setdeletedmarkup{\textcolor{blue}{\sout{#1}}}

\renewcommand{\added}[2][]{#2}
\renewcommand{\deleted}[2][]{}

\newcommand{\nDomainExperts}{n}
\newcommand{\nOnlineNarratives}{n}
\newcommand{\nPatientInterviews}{n}
\newcommand{\nClinicianInterviews}{n}
\newcommand{\nElicitedStories}{n}

\newcommand{\nPatientFormative}{n}

\newcommand{\nPatientEvaluation}{n}
\newcommand{\nClinicianEvaluation}{n}

\setlist[itemize]{noitemsep, topsep=0pt, parsep=0pt, partopsep=0pt}

\title{HealthTale: A Patient-Centric Health Story Visualization Tool}

\author{%
  \authororcid{Ryan Smith}{0009-0004-6304-8543}
  \authororcid{Kyle D. Chin}{0009-0001-0863-1101}
  \authororcid{Tamara Munzner}{0000-0002-3294-3869}
}

\authorfooter{
  \item
  	Ryan Smith, Kyle D. Chin and Tamara Munzner are with the Department of Computer Science, University of British Columbia. Email: \{rs0914,kdchin,tmm\}@cs.ubc.ca.
}

\abstract{Patients often struggle to communicate coherent accounts of their health histories during time-constrained clinical encounters. These accounts, which we refer to as health stories, include both clinical events and lived experiences. Existing systems prioritize structured, clinician-centered data and provide limited support for eliciting and communicating patient-generated narratives. We present HealthTale, a patient-centric visualization system designed to elicit health stories from patients and structure them to facilitate communication during initial clinical conversations. Its design arises from a multi-stage qualitative investigation across domain expert discussions ($\nDomainExperts=3$), online narratives ($\nOnlineNarratives=20$), patient ($\nPatientInterviews=11$) and clinician ($\nClinicianInterviews=6$) interviews, and elicited health stories ($\nElicitedStories=22$), identifying recurring patterns in how individuals construct and communicate their health stories. HealthTale transforms freeform narratives into structured timeline representations, grounded in a data abstraction that models health stories as events grouped by health concern and time. The abstraction captures both clinical and contextual information while accommodating temporally imprecise data and non-linear distributions of events across time. Through evaluation with patients ($\nPatientEvaluation=34$) and clinicians ($\nClinicianEvaluation=3$), we find that HealthTale supports recall, organization, and self-advocacy, while enabling clinicians to rapidly interpret patient-generated narratives and establish a shared understanding.
}


\teaser{
  \centering
  \includegraphics[width=\linewidth]{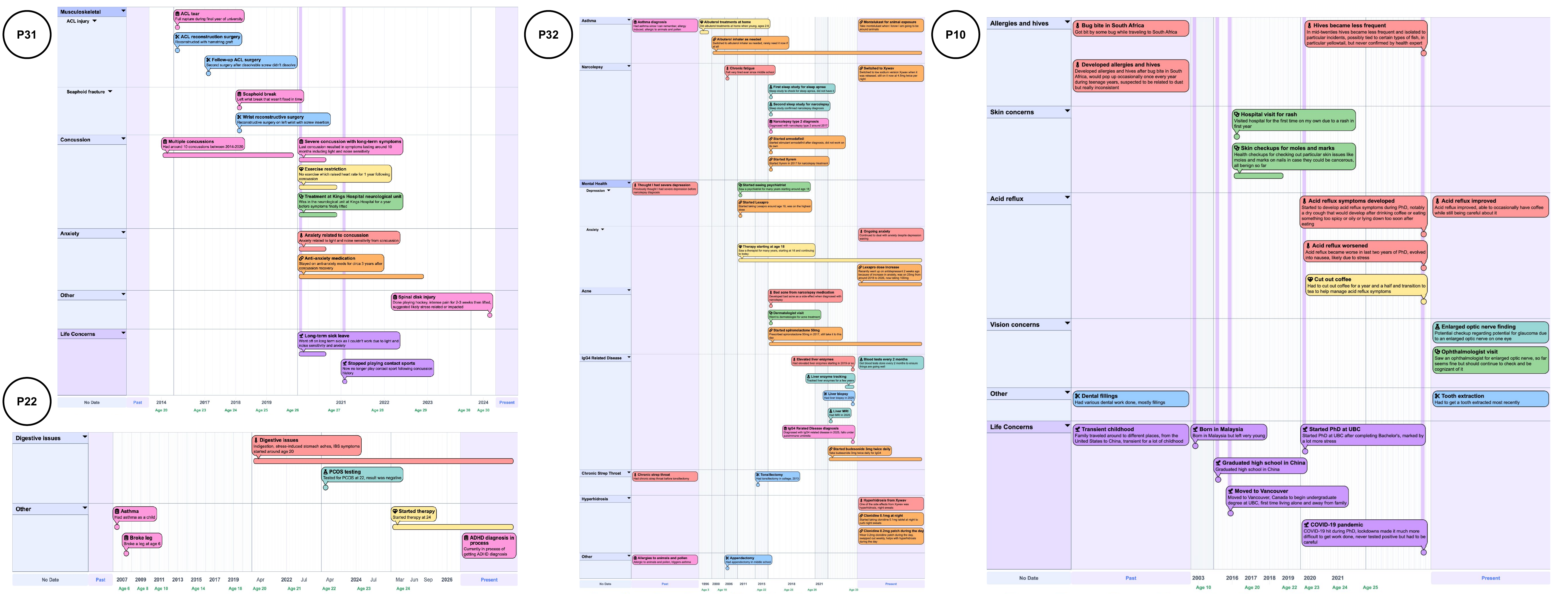}
  \caption{
  Example health stories created during the evaluation of the patient-centric HealthTale system. These visualizations illustrate the diversity of patient-generated narratives, ranging from relatively simple stories focused on a small number of Events to more complex stories spanning multiple Conditions and Timescales. HealthTale preserves individual differences in what patients choose to include and emphasize, and supports the integration of non-medical Life Concerns relevant to health outcomes, while providing an interpretable common overall structure to support rapid communication during time-constrained clinical encounters.  
  }

  \label{fig:healthstories}
}
\graphicspath{{figs/}{figures/}{pictures/}{images/}{./}} % where to search for the images

\usepackage{tabu}                      % only used for the table example
\usepackage{booktabs}                  % only used for the table example
\usepackage{lipsum}                    % used to generate placeholder text
\usepackage{mwe}                       % used to generate placeholder figures
\usepackage{ccicons}                   % package to be able to use icons from creative commons

\usepackage{mathptmx}                  % use matching math font
\begin{document}
\maketitle
%%%%%%%%%%%%%%%%%%%%%%%%%%%%%%%%%%%%%%%%%%%%%%%%%%%%%%%%%%%%%%%%
%%%%%%%%%%%%%%%%%%%%%% START OF THE PAPER %%%%%%%%%%%%%%%%%%%%%%
%%%%%%%%%%%%%%%%%%%%%%%%%%%%%%%%%%%%%%%%%%%%%%%%%%%%%%%%%%%%%%%%
%% The ``\maketitle'' command must be the first command after the
%% ``\begin{document}'' command. It prepares and prints the title block.
%% the only exception to this rule is the \firstsection command

\section{Introduction}

Patients must often convey complex, personal accounts of their health to clinicians in order to receive effective care, particularly during initial clinical encounters where little prior context is available. In these settings, clinicians rely heavily on patient-provided narratives to form an initial understanding of the patient’s condition, history, and priorities \cite{charon2001narrative}. These accounts extend beyond clinical facts to include contextual factors and personal interpretations that shape how individuals understand their health. We refer to these accounts as \textbf{health stories}: patient-generated narratives that integrate medical events with lived experience over time. While these narratives are central to clinical understanding, existing electronic medical record systems primarily capture structured, clinician-centered data, creating a gap between how patients understand their health and how it is represented in clinical contexts \cite{hickmann2022all, hall2011understanding, rajabiyazdi2021communicating, faisal2013making}.

We address this problem through a visualization system to elicit health stories from patients and help them communicate those stories with clinicians, with a particular focus on supporting initial clinical visits where patients must efficiently convey their history to new providers. The system was designed in collaboration with a project partner, a digital health solutions company and its clinical team.  

We began by conducting a multi-stage qualitative investigation into patient-generated narratives via discussions with clinical team domain experts ($\nDomainExperts=3$), collecting and analyzing online health stories ($\nOnlineNarratives=20$), interviews with patients ($\nPatientInterviews=11$) and clinicians ($\nClinicianInterviews=6$), and directly eliciting written health stories ($\nElicitedStories=22$). Through this process, we identified recurring patterns in how patients think about and communicate their health stories, especially in the context of preparing for and navigating first-time clinical interactions.

From these observations, we derive a data abstraction that models health stories as a series of events with associated attributes and organizes them across two cross-cutting grouping structures: the type of health concern, and the timing of when they occurred. This abstraction captures both clinical information and contextual experience, while supporting events specified with a broad range of temporal precision and that are unevenly distributed across time. 

We present \textbf{HealthTale}, a patient-centric health story visualization system designed to support this data abstraction by transforming freeform narratives into structured timeline representations. The system supports patients in preparing for initial clinical encounters and communicating with clinicians to help them rapidly form a coherent understanding of a patient’s history. We evaluate HealthTale through studies with patients ($\nPatientEvaluation=34$) and clinicians ($\nClinicianEvaluation=3$), showing that it supports patient recall, organization, and self-advocacy, while providing preliminary evidence that these representations enable clinicians to quickly interpret patient-generated health stories during early-stage interactions.

The primary contributions of this work are:
\begin{itemize}
    \item A flexible data abstraction for health stories, grounded in a multi-stage qualitative investigation of patient-generated narratives across diverse sources.
    \item HealthTale, a patient-centric Health Story visualization system that transforms elicited free-form text into this structured data abstraction and visually encodes it to support patient-clinician communication, particularly for initial clinical encounters. We provide preliminary evidence of utility through evaluation with patients and clinicians, demonstrating support for both health story elicitation and communication.
\end{itemize}

The secondary contributions of this work are:
\begin{itemize}
    \item A corpus of health stories ($n=85$) collected across multiple sources.
    \item A grouping and layout algorithm for visualizing health stories that accommodates events specified at different levels of temporal precision, and with non-linear temporal distribution.
\end{itemize}

\section{Background and Motivation}

Patient-centered care emphasizes understanding a patient’s experiences, context, and priorities when making clinical decisions \cite{hickmann2022all}. When patients feel heard and understood, health outcomes improve. Increased participation has been linked to greater engagement with treatment, improved understanding of health status, and increased empowerment in managing conditions \cite{higgins2017unraveling, zhu2017sharing, bhattacharyya2019using}. At a system level, effective patient–clinician communication is also associated with improved healthcare efficiency and reduced costs \cite{hickmann2022all, street2009does, stewart1995effective, zachariae2003association, zolnierek2009physician}.

In practice, however, communicating these narratives remains challenging, particularly during initial clinical visits where clinicians have little prior context and patients must establish their health story from scratch \cite{kong2020behavioral, ryu2023connections}. Unlike follow-up encounters, where shared understanding can be incrementally refined, first-time interactions require patients to summarize complex and sometimes long-term histories to a new clinician within a constrained time window. During these encounters, patients typically complete structured intake forms in advance and then verbally summarize their histories within the limited first few minutes of an appointment \cite{norouzinia2015communication}. Intake forms prioritize standardized, clinician-centered data and provide little space for contextual detail, while verbal accounts are often incomplete, non-linear, and difficult to organize. As a result, clinicians must reconstruct patient histories from fragmented information, meaning important contextual details may be overlooked \cite{sepehri2023beyond, hickmann2022all, rajabiyazdi2017differences, schaad2015dissatisfaction}.

Existing tools do not adequately support this process. Electronic medical records (EMRs) provide structured representations of patient data but are fundamentally clinician-centered, focusing on coded medical events such as diagnoses, treatments, and tests \cite{hudson2025outcomes,esmail2021evaluation}. While effective for documentation and clinical review, they omit the contextual and experiential elements that patients use to make sense of their health \cite{faisal2013making,rajabiyazdi2021communicating}. Similarly, patient-facing tools often rely on predefined inputs or continuous tracking, capturing discrete data points rather than supporting patients in constructing and organizing their own health stories, and can introduce cognitive and emotional barriers to effective communication \cite{kong2020behavioral}. A disconnect remains between patient-generated health stories and their representation within clinical systems.

This gap motivates two core design goals:

\begin{itemize}
  \item \textbf{DG-E, Elicitation:} Support patients in constructing relevant and context-rich health stories, enabling recall and self-advocacy.
  
  \item \textbf{DG-C, Communication:} Support efficient communication through a static, shareable visualization artifact brought by the patient to an initial encounter with a clinician, enabling the rapid formation of a shared mental model. 
\end{itemize}

These goals aim to improve both the efficiency and quality of early-stage clinical interactions by aligning patient-generated narratives with clinically interpretable representations.

\section{Related Work}

We situate our work within prior research on eliciting patient information, electronic medical record visualization, and narrative visualization. 
%Rather than cataloging systems, we highlight evidence and limitations that motivate the need to support patient-generated health story elicitation.

\subsection{Eliciting Patient Information}

Healthcare systems commonly rely on structured methods such as intake forms, symptom checklists, Patient-Reported Outcome Measures (PROMs), and Patient-Reported Experience Measures (PREMs) to collect patient information \cite{hudson2025outcomes, manalili2021using, esmail2021evaluation, valenstein2008formatting, leslie1994standardization}. These approaches improve consistency and efficiency but constrain how patients express their experiences, often limiting responses to predefined categories.

Digital tools, including electronic forms, health applications, and wearables, extend these approaches but continue to prioritize structured data collection over patient-generated narratives \cite{lagan2020digital, smuck2021emerging, canali2022challenges}. As a result, they often fail to capture contextual and experiential information that patients consider important \cite{campbell2022perceived, dodson2024capturing, gilmore2023uses}. Additionally, these systems frequently require consistent data entry, which can be burdensome and difficult to maintain over time \cite{rapp2016personal, smith2007integrating, kientz2009baby, herrmann1995reporting, choe2014understanding}.

Narrative approaches allow patients to describe their experiences in their own terms and highlight meaningful relationships between events \cite{detmar2000you, rajabiyazdi2021communicating}. Such narrative accounts have also been shown to support reflection and personal understanding of one’s health over time \cite{li2011understanding}. However, such narratives are difficult to structure, compare, and integrate into clinical workflows. This tension between flexibility and structure motivates the need for approaches that can transform patient-generated narratives into interpretable representations.

Our work addresses this gap by enabling freeform narrative input while automatically deriving structured representations for visualization, preserving expressiveness while improving interpretability and communication in clinical contexts.

\subsection{Electronic Medical Record Visualization}

Visualization systems for electronic medical records (EMRs) commonly represent patient histories as timelines of diagnoses, treatments, and clinical Events \cite{plaisant1996lifelines, wongsuphasawat2011lifeflow, hirsch2015harvest, rostamzadeh2020data}. These systems support clinicians in navigating large volumes of structured data and identifying temporal patterns \cite{bade2004connecting}.

However, EMR visualizations are inherently constrained by the structure and scope of the underlying data \cite{plaisant1996lifelines}. They focus on clinically coded events and omit contextual and narrative elements present in patient-generated health stories. As a result, they reflect a clinician-centered view of patient history and do not support how patients construct or communicate their experiences. The intersection between EMR visualization and narrative-driven representations remains limited, highlighting an opportunity for approaches that incorporate patient-generated data \cite{sultanum2022chartwalk, rajabiyazdi2021communicating}.

This gap motivates our work, which bridges clinician-centered EMR visualizations and patient-authored narratives by transforming health stories into structured, interpretable visual representations.

\subsection{Visualization for Storytelling}

Narrative visualization explores how visual representations can structure and communicate stories by guiding interpretation and emphasizing relationships between events \cite{segel2010narrative, figueiras2014narrative}. Such representations can support comprehension and memorability, reduce cognitive load, and improve understanding of temporal relationships \cite{kim2017explaining}, while enabling subjective organization of narratives \cite{carpendale2017subjectivity}.

Some visualization systems extend this work by enabling users to construct visualizations through graphical interfaces, supporting non-expert users in creating and externalizing their own representations \cite{Kim2019DataToon, Kim2019Inking}. Systems such as timeline-based storytelling tools allow users to flexibly structure temporal narratives \cite{Offenwanger2024TimeSplines, fulda2016timeline}, while others support reflection over personal data \cite{Liang2016SleepExplorer, li2011understanding, huang2014personal}.

However, none of this prior work addresses the specific needs of patient-driven communication of a health story during a clinical encounter. 

Moreover, prior work on storytelling assumes that the underlying data is already structured or requires users to manually organize it, so these systems do not address how unstructured, freeform narratives can be transformed into consistent representations that support both elicitation and communicating. In particular, they do not account for the variability, ambiguity, and temporal imprecision inherent in patient-generated health stories, limiting their effectiveness in clinical contexts.

All of these past approaches demonstrate the potential of narrative visualization, but it remains largely unexplored in the context of patient-generated health stories. Our work addresses this gap by applying narrative visualization techniques to patient-authored health stories, transforming freeform narratives into structured, interpretable representations for clinical communication.

\section{Data Abstraction}

We both derive a data abstraction for health stories and motivate our two design goals through a multi-stage data collection and analysis process involving healthcare domain experts, patients, and clinicians. \deleted{Across these stages, we examine how individuals construct, group, and temporally describe their health experiences in narrative form.} From this analysis, we identify recurring structural patterns and use them to inform a formal representation of health stories and the design goals that guide system development. We first present the formalization of this representation, followed by an explanation of the process and the empirical observations resulting from it that motivated and justified our choices. 

\subsection{Data Abstraction Formalization}

\begin{figure}[!t]
  \centering
  \includegraphics[width=0.8\columnwidth]{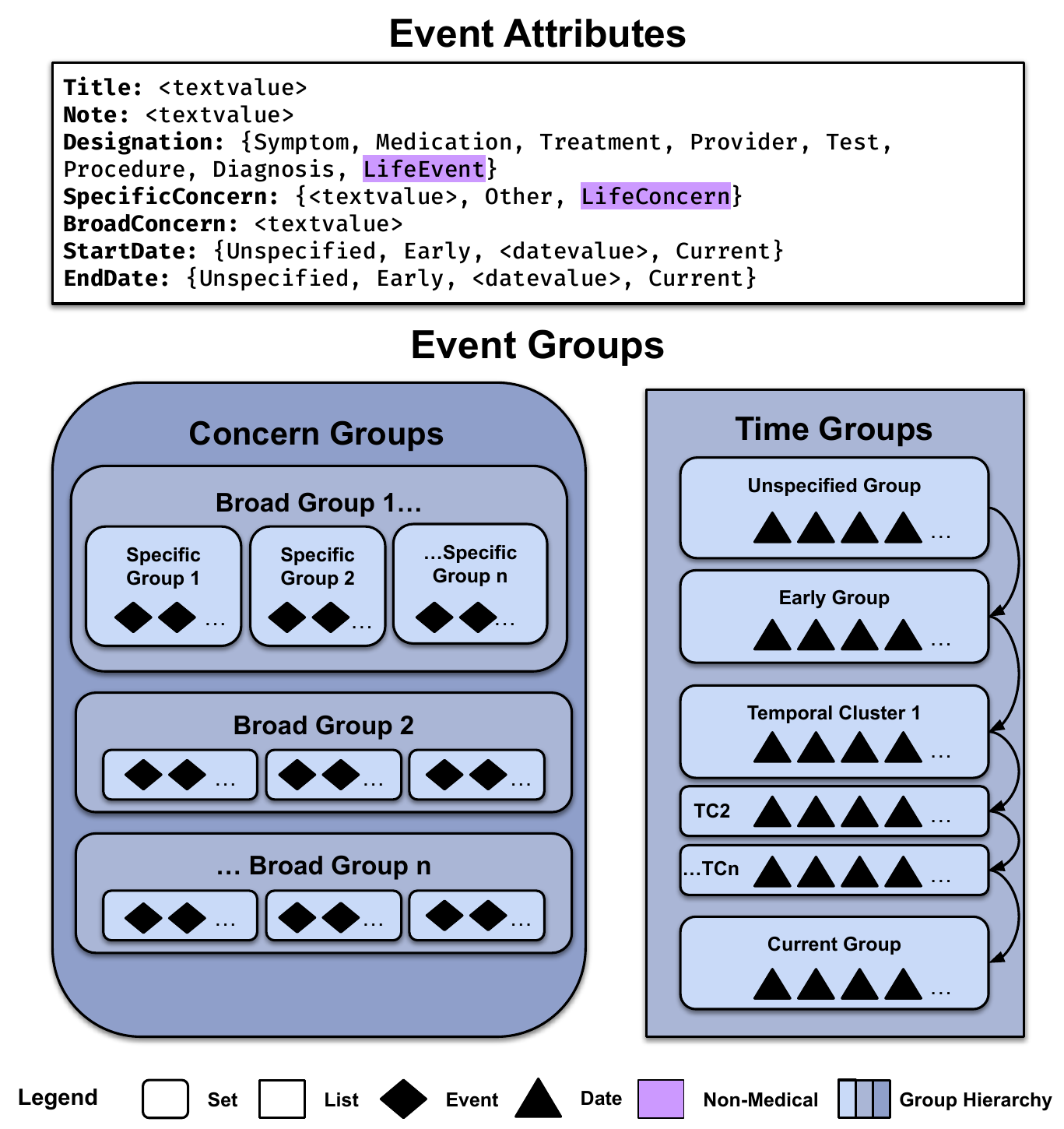}
  \caption{
    Health story data abstraction. Events are atomic items with multiple attributes. Some attributes, such as \textit{Title}, have freeform text as values, while others, like \textit{Designation}, use predefined values; some, such as \textit{Concern}, support both. \textit{Life Event} and \textit{Life Concern} are highlighted in purple to emphasize that these non-medical elements are typically absent from standard medical abstractions. Events are organized into two cross-cutting structures: \textit{Concern Groups} and \textit{Time Groups}.
  }
  \label{fig:dataabstraction}
\end{figure}

Fig.~\ref{fig:dataabstraction} presents our data abstraction, which centers an \textbf{Event} as the atomic unit of a health story. Events in a health story represent changes, processes, or interactions within an individual’s health experience, such as the onset of a symptom, the initiation of taking a medication, a specific clinical visit, or a non-medical incident or experience that pertains to health outcomes. 

\subsubsection{Event Attributes}

Each Event contains a set of attributes that describe what occurred, what it relates to, and when it happened. The attributes associated with each Event were derived from recurring patterns observed in how participants described and structured these experiences, so as to %These attributes reflect both clinical information and lived experience, 
capture the variability and ambiguity inherent in patient-generated narratives.

The \textbf{Title} and \textbf{Notes} attributes have values that are free-form text; Title captures the primary descriptor of the Event, like the name of a medication or a very brief summary of what occurred, while Notes provide additional contextual detail. \deleted{This separation between title and supporting information supports expressive input while preserving structured interpretation.}

\textbf{Designation} is a categorical attribute describing the type of Event, with eight possible values: \textit{Symptom}, \textit{Medication}, \textit{Treatment}, \textit{Provider}, \textit{Test}, \textit{Procedure}, \textit{Diagnosis}, and \textit{Life Event}. The first seven correspond to standard clinical categories, while \textit{Life Event} captures non-medical experiences like job changes or moves or anything that is central to how patients understand and communicate their health. This category is not typically captured in structured medical records.  

The \textbf{SpecificConcern} and \textbf{BroadConcern} attributes describe what kind of problem the Event is associated with. SpecificConcern identifies a particular problem, like \textit{arthritis}, with a free-form text value. It may also be set to the pre-specified values \textit{Other}, indicating that there is no specific associated problem. For all Events of type Life Event, this attribute is set to the pre-specified value of \textit{Life Concern}, the category for all non-medical problems. BroadConcern represents a higher-level grouping, such as \textit{musculoskeletal} or \textit{mental health}, with a free-form text value, and is optional. These attributes are used to create groups, as described below.
%Together, these attributes support flexible grouping while preserving both medical and contextual relevance.

\textbf{StartDate} and \textbf{EndDate} describe when an Event occurs. Each attribute takes one of four values: \textit{Unspecified}, \textit{Early}, \textit{Date}, or \textit{Current}. \textit{Unspecified} indicates that no temporal information is available for that aspect of the Event. \textit{Early} indicates that no explicit date is given, but the surrounding health story provides enough information to infer that the Event occurred before the first Event with explicitly specified timing. \textit{Date} represents an explicitly provided temporal reference by the patient, which may be specified either as an absolute calendar date, or as a relative age. Relative age values are ultimately mapped into calendar dates, using the date of birth. \textit{Current} indicates that no explicit date is given, but there is implicit meaning that the Event is ongoing at the present time. \added{For example, ongoing routines can be represented as Current Events when patients frame them as part of the story.}
%Absolute calendar dates and Relative age provide fully specified times, while Current captures ongoing Events, Early captures Events occurring prior to the first fully specified reference point, and Unspecified represents Events with no temporal information. 
This formulation explicitly supports the imprecision and variability observed in patient narratives. Events may be loosely situated in time or entirely without timestamps, enabling patients to construct meaningful representations without conforming to the rigid temporal requirements typical of EMRs.

\subsubsection{Event Groups}

Events are organized across two cross-cutting grouping structures: Concern Groups and Time Groups. These structures are derived from attributes across the full set of Events and provide complementary organizational views of that set.

%\paragraph{Concern Groups}
\textbf{Concern Groups} are structured as a two-level hierarchy of sets: a SpecificGroup is a set of Events, and a BroadGroup is a set of SpecificGroups. Both a \textbf{SpecificGroup} and a \textbf{BroadGroup} simply contain all Events with the same values for attributes of SpecificConcern and BroadConcern respectively, and their name is set to that value. 
%This structure enables Events to be organized by problem while preserving flexibility. The inclusion of \textit{Life Concern} ensures that non-medical context is explicitly represented rather than excluded, while \textit{Other} allows Events to remain uncategorized when no clear condition applies. This structure reflects how patients describe their experiences, which often extend beyond clearly defined clinical categories.

%\paragraph{Time Groups}
\textbf{Time Groups} operate on the individual StartDate and EndDate attributes rather than on whole Events. Each date associated with an Event is assigned independently to a Time Group based on its value, meaning a single Event may be associated with multiple Time Groups if its StartDate and EndDate fall into different partitions.

Time Groups are structured as a list containing sets of dates, partitioned according to the \textbf{StartDate} and \textbf{EndDate} attribute values. The \textbf{Unspecified}, \textbf{Early}, and \textbf{Current} Groups are sets that contain all dates with those respective values. Events with explicit date values are organized into one or more \textbf{Temporal Clusters}, each of which is a set of dates derived from global patterns in the data, allowing the non-uniform temporal distribution of health stories to be explicitly represented.

%\subsubsection{Cross-Cutting Structure}

\deleted{The combination of Concern Groups and Time Groups provides a cross-cutting structure to capture the salient aspects of patient narratives in a very flexible way.}
%Events are organized simultaneously across Concern and Time Groups, with Event attributes determining how they are interpreted within each structure. This enables multiple valid views over the same underlying data and reflects how patients describe their histories: in terms of what happened, what it relates to, and when it occurred, often in overlapping and interdependent ways.
\deleted{Unlike traditional EMR data abstractions, which prioritize precise, standardized, and clinician-centered representations, this model accommodates temporal ambiguity and integrates non-medical contextual information.}
%By separating Event attributes from grouping structures, the abstraction supports flexible organization while preserving the expressiveness of patient-generated narratives.

\subsection{Data Abstraction Process and Evolution}

Our data abstraction and the articulation of our design goals emerged from 
%To both derive our data abstraction and motivate our design goals, 
%We conducted 
a multi-stage data collection and analysis process centered on gathering and analyzing health story artifacts, \added{triangulating} across different contexts. 
\deleted{This analysis examined how health stories are constructed, communicated, and interpreted, allowing us to iteratively refine both the abstraction and the design goals for supporting health story elicitation and communication.} \added{The first author conducted the qualitative coding at each stage, with iterative refinement arising from cross-checking initial findings against artifacts created at later stages, and through discussion with the entire research team and clinical collaborators.}

All studies with human participants described in this paper were approved by our university’s Behavioural Research Ethics Board(H23-02636), and we obtained informed consent from all participants.

\subsubsection{Initial Domain Characterization (Expert Discussion)}

We began with an initial domain characterization phase to ground our understanding of health stories within clinical practice. We conducted dozens of interactions with our industry partner’s internal clinical team ($\nDomainExperts=3$), all of whom are practicing healthcare professionals. These interactions varied in length and modality, ranging from brief Slack discussions to formal 30--60 minute meetings.

Through these discussions, we developed a foundational understanding of the clinical context surrounding patient intake and how patient histories are interpreted in practice. The clinicians emphasized the importance of key diagnoses and treatments, and their temporal progression, when forming an understanding of a patient’s health, particularly in the early moments of an encounter. These insights informed which types of health information should be represented in our system and directly influenced the development of the Designation attribute in our abstraction, supporting both the capture of relevant patient information for DG-E (Elicitation) and its efficient interpretation in clinical settings for DG-C (Communication).

Following this initial domain characterization, we refined the scope of the problem and shifted our focus toward understanding how people initially tell their health stories in practice. To do this, we collected health story artifacts across multiple contexts to examine how these narratives are constructed, communicated, and externalized by real patients. We designed a multi-stage data collection process spanning natural, clinical, and controlled settings, allowing us to identify recurring structures and inform both the data abstraction and design goals.

\subsubsection{Stage 1: Unsolicited Health Stories (Online Narratives)}

\textbf{Motivation.}
We first sought to understand what constitutes a health story by examining naturally occurring, unsolicited patient narratives, using online posts as a proxy for how individuals externalize and communicate their experiences outside of clinical settings.

\noindent
\textbf{Methods.}
We analyzed online health stories, using a netnographic approach \cite{kozinets2020netnography} that
adapts ethnographic methods to digital contexts. We selected posts ($n=20$) from the \texttt{/r/chronicillness} subreddit that represented apparent individual health stories. \added{We included first-person health stories and excluded advice requests or isolated symptom questions.} \deleted{These unsolicited narratives provide insight into how patients describe their experiences without external prompting.}

The first author performed \added{thematic analysis}~\cite{braun2006using} and iteratively open coded the stories, decomposing each into individual Events and developing a coding schema of Designations through repeated passes over the data. Coding proceeded until termination, defined as the point at which no new codes were identified and the existing schema sufficiently covered new stories.

\noindent
\textbf{Data Abstraction Results.}
We observed that Events may include temporal information as either singular instances or ranges, and that patients employ diverse temporal strategies. Narratives often reflect a dominant temporal framing, but may interleave multiple strategies or omit temporal information entirely. These observations motivated a flexible StartDate and EndDate representation, including support for unspecified cases. We also observed variation in the granularity and density of Events over time, motivating support for having multiple Time Clusters.

\noindent
\textbf{Design Goal Results.}
These findings support DG-E by highlighting the need to capture diverse, patient-driven narrative structures, including incomplete and non-standard temporal information. They also support DG-C by motivating representations that preserve temporal variation and density, enabling clinicians to interpret patterns of activity and inactivity over time. 

\subsubsection{Stage 2: Patient and Clinician Interviews}

\textbf{Motivation.}
We next examined how health stories are constructed and communicated in clinical contexts, focusing on how patient narratives align with clinical needs.

\noindent
\textbf{Methods.}
We conducted semi-structured interviews with patients ($\nPatientInterviews=11$; age range = 24--69; $M = 49.1$, $Mdn = 54.5$) and clinicians ($\nClinicianInterviews=6$; age range = 31--43; $M = 38.8$, $Mdn = 39.5$; years in field range = 3--19; $M = 10.0$), including registered nurses in internal medicine and critical care, a speech pathologist, a registered physiotherapist, and an associate professor of clinical medicine. Interviews combined roleplaying scenarios with process-oriented discussion of how health stories are shared and interpreted. \added{Interviews lasted approximately 60 minutes and included first-visit roleplay and questions about story organization.}

We recruited patients with complex or chronic conditions from our project partner’s network (CAD\$25 honorarium) and clinicians from the same network (CAD\$40 honorarium). 

The first author performed thematic analysis and iteratively open coded interview transcripts, analyzing patient and clinician responses as separate but related corpora to capture differences in how health stories are produced and interpreted. We reached termination when additional interviews no longer produced new codes and instead revealed limitations of spoken recall, motivating a shift toward eliciting health stories in a more structured, written form.

\noindent
\textbf{Data Abstraction Results.}
We identified systematic differences between clinical expectations and patient-generated narratives, reinforcing the need for a representation that bridges these perspectives. Patients described Events using varied and often imprecise temporal strategies, including missing, relative, and ongoing time references. These observations further supported the StartDate and EndDate Unspecified, Early, and Current values.

\noindent
\textbf{Design Goal Results.}
These findings support DG-E by highlighting challenges in recall, ordering, and completeness when patients describe their histories verbally, motivating support for structured yet flexible narrative construction. They support DG-C by emphasizing the time-constrained nature of clinical encounters and the need for concise, structured, and easily scannable representations. Participants described bringing prepared materials into visits and noted the difficulty of communicating with clinicians through long, unstructured narratives, motivating representations that align patient expression with clinical interpretability while preserving patient perspective. Although patients commonly relied on printed materials, the clinicians emphasized the need for electronically shareable artifacts. This finding elaborated our understanding of DG-C with the idea that the end result of using the system should be an artifact that remains interpretable without interaction, supporting effective communication within static, artifact-based clinical workflows.

\begin{figure*}[t]
  \centering
  \includegraphics[width=.85\textwidth]{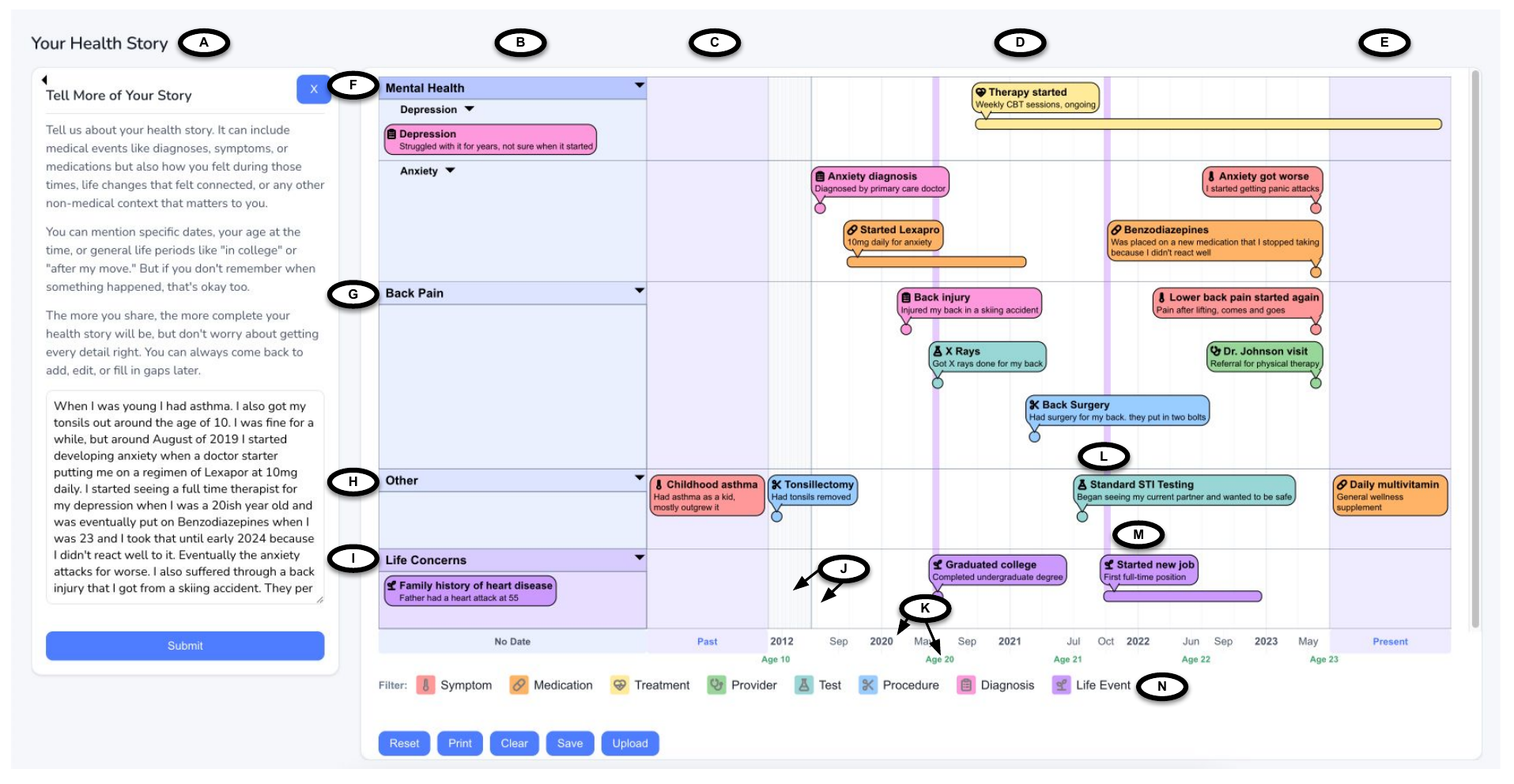}
  \caption{
%HealthTale provides structure and organization to visually represent complex medical histories as health stories told from the patient’s perspective. The patient writes free-form text in the Elicitation Panel (A), which is parsed according to our three-part Event data abstraction of Concern, Designation, and Time Specification and visually encoded in the canvas. The canvas is organized into temporal segments, including the No Time segment (B), Past segment (C), series of Timescale segments (D), and Present segment (E). Events are distributed across horizontal tracks based on their associated Concerns, including grouped Condition tracks (F), individual Condition tracks (G), an Other track (H), and a Life Concerns track (I). Temporal granularity within the timeline is supported by multiple Timescales and grid lines (J), while the horizontal axis includes both absolute date labels and relative age labels (K). Individual Events are represented as InfoBoxes (L), with Life Events additionally indicated by vertical lines spanning tracks to support cross-track comparison (M). Event color encodes Designation (N).
In HealthTale, after the patient writes free-form text in the Elicitation Panel on the left (A), it is automatically transformed into structured data visually encoded in the canvas on the right. The vertical segments show temporally ordered groups, starting with No Time (B), then Past (C), with multiple Timescale segments (D), followed by Present (E). Events are contained within horizontal tracks based on Concerns, with two-level groups of Broad and Specific Conditions (F), followed by individual Condition tracks (G), with Other (H) and Life Concerns (I) at the bottom. The density of the multiple Timescales are shown with grid lines (J), with absolute date labels and relative age labels on the horizontal axis (K). Individual Events are represented as InfoBoxes (L), and the Life Events also have vertical lines spanning tracks to support cross-track comparison (M). Event color encodes Designation (N).
  }
  \label{fig:interface}
\end{figure*}

\subsubsection{Stage 3: Elicited Health Stories (In-Person Sessions)}

\textbf{Motivation.}
Finally, we sought to understand how patients construct health stories when explicitly asked to externalize them in written form.

\noindent
\textbf{Methods.}
We conducted an in-person elicitation study with patients ($\nElicitedStories=22$; age range = 19--71; $M = 32.4$, $Mdn = 28.5$), where each session lasted approximately 60 minutes. Participants were recruited as individuals with complex health conditions through our project partner’s network and via flyers distributed across the city, and received a CAD\$30 honorarium.

Participants were asked to create written health story artifacts in their own words, followed by a reflection on their thought process. The first author, again, performed thematic analysis and iteratively open coded these artifacts, building on and refining the code structures developed in Stages 1 and 2. Coding continued until additional artifacts did not yield new codes, indicating that the existing schema sufficiently captured the observed structures.

\noindent
\textbf{Data Abstraction Results.}
We observed that patients frequently include contextual and non-medical experiences, leading us to formalize Life Events as a distinct Designation. Patients often described distant Events using vague temporal references and explicitly identified ongoing Events, reinforcing the need for Early and Current values. We also observed that Events cluster within specific periods, with bursts of activity and gaps, motivating support for multiple Temporal Clusters.

\noindent
\textbf{Design Goal Results.}
These findings support DG-E by showing that patients naturally construct narratives that combine freeform expression with implicit structure, motivating systems that support flexible input while capturing underlying organization. They also support DG-C by highlighting that patients prioritize certain Events while omitting others they consider routine, suggesting the need for representations that preserve patient priorities while enabling clinicians to quickly identify relevant gaps and patterns.

\section{HealthTale Design}

HealthTale is a patient-centric health story visualization system 
%designed to support both elicitation, DG-E, and communication, DG-C. 
%The system operationalizes our data abstraction by transforming freeform text into an interpretable visual representation of structured Events with Designation, Concern, and Time attributes, organized into cross-cutting Concern Groups and Time Groups.
%To support \textbf{DG-E}, HealthTale enables low-burden narrative input via narrative text that is automatically structured into meaningful representations. To support \textbf{DG-C}, it produces structured and interpretable static visual artifacts that facilitate rapid understanding during time-constrained clinical encounters. These components are 
with a unified interface that couples elicitation through free-form text (DG-E), automated structuring of that text according to our data abstraction, and its visual encoding designed for efficient communication with a clinician (DG-C). \added{The intended workflow is for patients to use HealthTale in advance of an initial clinical encounter by typing their health story into it to generate a visualization, then bring a printout or digital copy of that visualization to that visit to support their ommunication with the clinician.}

%\subsection{Overview and User Flow}

Fig.~\ref{fig:interface} shows the HealthTale interface, which has a collapsible information pane on the left and a visualization canvas on the right. 
%The information pane displays one of three panels at a time: 
%the Health Profile Panel \added{for name and date of birth}, the Elicitation Panel \added{for freeform story entry}, or the Event Editing Panel \added{for manual addition and correction of Event attributes}.
The user flow of HealthTale begins with the Health Profile Panel \added{visible in the information pane on the left}, where users provide their name and date of birth, which enables relative time expressions to be resolved into absolute time. \added{The left information pane then updates to show the} Elicitation Panel \deleted{then appears}, with lightweight prompts asking \added{users}\deleted{them} to type their health story as freeform text (Fig.~\ref{fig:interface}(A)).
This input is automatically transformed into a structured visualization in the canvas \added{on the right}. Users can refine this representation if desired, by adding new events and editing existing Event attributes in the Event Editing Panel \added{on the left}. \deleted{The resulting visualization can be saved and printed as a static artifact for use during clinical encounters.} 

\subsection{Visual Encoding}

The visual encoding of the structured health story positions \deleted{InfoBoxes representing Events into} \added{Events within} horizontal Tracks according to Concerns and vertical Segments ordered by Time. 
%The visualization canvas presents the structured health story as a timeline-based representation.

\subsubsection{Event InfoBoxes}

Each Event item is encoded with an \textbf{InfoBox} glyph (Fig.~\ref{fig:interface}(L)), which serves as the primary visual unit of the health story. Each InfoBox includes the Title and Note text, placed within a surrounding box with rounded corners that is sized to encompass both fields. The horizontal position of InfoBoxes is determined by the Event's StartDate and EndDate attributes and their vertical position by the Concern. 

Temporal extent is encoded using a marker placed directly below the main box, where a circle encodes a single point in time and a horizontal line encodes a duration. The box is styled to look like a dialogue bubble, with the downward triangle pointing to the marker. 
The box and marker are color coded according to the Designation attribute (Fig.~\ref{fig:interface}(N)), which is also encoded with an icon in the upper left. \added{The eight-category color scheme follows our project partner’s established conventions.}

\subsubsection{Concern Group Horizontal Tracks}

Concern Groups are shown as horizontal \textbf{Tracks} that contain Events, vertically ordered with the larger two-layer BroadCondition sets (Fig.~\ref{fig:interface}(F)) at the top and the singleton SpecificCondition sets below (Fig.~\ref{fig:interface}(G)). Within these categories the positioning follows the narrative order in the health story, except that the \emph{Other} track (Fig.~\ref{fig:interface}(H)) is always just above the \emph{LifeConcern} track at the very bottom, if these groups exist (Fig.~\ref{fig:interface}(I)). The InfoBoxes for Life Events are augmented with a vertical purple line spanning tracks (Fig.~\ref{fig:interface}(M)), to make temporal correspondences between these and all other Events easy to notice. 

%within which the Events in those sets are contained. A BroadGroup set (Fig.~\ref{fig:interface}F) is demarcated with a dark blue box around its name on the left side, with its constituent SpecificGroup (Fig.~\ref{fig:interface}G) sets nested within it and their names indented. A SpecificGroup set with no enclosing BroadGroup set has no name indentation. We also include an \emph{Other} track (Fig.~\ref{fig:interface}H) for Events with the corresponding SpecificConcern value. If the LifeConcern track exists (Fig.~\ref{fig:interface}I), it always appears at the bottom, and Life Event InfoBoxes are augmented with a vertical purple line spanning tracks (Fig.~\ref{fig:interface}M), emphasizing their cross-cutting role.

\subsubsection{Time Group Vertical Segments}

The ordered list of Time Groups are shown as vertical \textbf{Segments}. On the left is the Unspecified Group labelled as \emph{No Time} (Fig.~\ref{fig:interface}(B)), followed by the Early Group labelled as \emph{Past} (Fig.~\ref{fig:interface}(C)), and the rightmost segment contains the Current Group labelled as \emph{Present} (Fig.~\ref{fig:interface}(E)). These segments do not display explicit axes. 

\added{Health stories often contain dense bursts of Events separated by long inactive periods, so non-uniform Timescales condense these gaps and preserve readable space for information-rich periods.} Each Temporal Cluster is represented within the absolute timeline as a \textbf{Timescale} (Fig.~\ref{fig:interface}(D)), with boundaries indicated by darker separators. Within each Timescale, the density of the grid lines (Fig.~\ref{fig:interface}(J)) conveys the temporal density of each segment.

Although absolute calendar time and relative age date values are unified within the data abstraction, they are represented distinctly in the visual encoding to preserve the nuance in how patients express time in their health stories. The horizontal axis uses dual labels to show complementary temporal context, with absolute encoded above in black and relative below in green (Fig.~\ref{fig:interface}(K)).

\subsection{Algorithm and Architecture}

HealthTale translates freeform health stories into structured visualizations through a pipeline consisting of three stages. First, the text is parsed to create a structured representation based on Events. Then, a grouping and layout algorithm creates Event Groups, determines the size of Tracks and Segments, and arranges the Event InfoBoxes within them. Finally, a visual styling pass fine-tunes the appearance of all the on-screen elements. 
%Together, these components operationalize the data abstraction to produce an interpretable visualization that supports both elicitation and communication.

\subsubsection{Parsing}

Given a freeform narrative, the parser identifies Events and assigns their attributes, extracting as much temporal information as is available to record in the Time attributes. \added{The prompt constrains output to the Event schema described above, so downstream layout receives Events with Designation, Concern, and StartDate/EndDate fields.}

We use a large language model (LLM), Claude Sonnet 4.6 \cite{anthropic2025claude}, to parse freeform health stories into structured data that incorporates all of the previously described attributes, along with a unique per-item identifier. The model is used strictly for transformation, not generation, and extracts Events and their associated attributes without introducing new information. The system is accessed via an API key tied to a version of the model approved by our project partner for handling medical data, ensuring that all processing aligns with appropriate data governance requirements.

\subsubsection{Grouping and Layout Algorithm}

The grouping and layout algorithm is designed to support health story representations for initial clinical encounters, where clinicians must rapidly interpret patient histories under time constraints. 
The design target thus is a moderate number of Events, typically on the order of dozens \added{but not hundreds}. We prioritize readability, compactness, and interpretability over scalability to arbitrarily large inputs. These assumptions inform key design decisions in the algorithm, including the use of adaptive temporal clustering, density-based width allocation, and layout optimizations that reduce canvas vertical height for artifact-based use. Fig.~\ref{fig:pseudocode} summarizes the algorithm as pseudocode.

\begin{figure}[t]
\begingroup
\color{black} 
\begin{Verbatim}[fontsize=\small,baselinestretch=0.70]
Timeline_Layout(events, W) { // Events array, window width in pixels
    clusters = Cluster_Temporal_Events(Temporal(events))
    bestLayout = null
    for each r in [10%, 15%, ..., 50%]: // ratio of widths
      layout = Draft_Layout(events, clusters, W, r)
      if bestLayout is null or
         layout.height < bestLayout.height:
        bestLayout = layout
    return bestLayout
}
Cluster_Temporal_Events(events) {
    spans = NormalizeDurations(events) // normalize to range 0-100
    eps = min(30, (2.5 yrs / TotalYears(events)) * 100)
    minPts = 1 // allow singleton clusters
    clusters = DBSCAN(spans, eps, minPts)
    return clusters
}
Draft_Layout(events, clusters, W, r) {
    segments = MakeSegmentsArray(events, clusters)
    Position_Segments(segments, W, r)
    tracks = Pack_Concern_Tracks(events, segments)
    height = TotalHeight(tracks)
    return segments, tracks, height
}
Position_Segments(segments, W, r) {
    x = W * r
    for each s in segments:
      s.width = (DateCount(s) / TotalDateCount(segments))
                * W * (1-r)
      s.scale = MapFromDatesToPixels(DateRange(s), s.width)
      s.left = x
      s.right = s.left + s.width
      x = s.right
}
Pack_Concern_Tracks(events, segments) {
    tracks = Group_By_Concern(events)
    h = 0
    for each track in tracks:
      for each event in track:
        infobox = CalcInfoBoxExtentsAndHorizPosition(
                    event, segments)
        lane = FirstAvailableLane(track, infobox)
        if !lane: lane = NewLane(track)
        PlaceInLane(lane, infobox)
        IncreaseLaneHeightIfNecessary(lane)
      track.height = SumLaneHeights(track)
      track.top = h
      track.bottom = track.top - track.height
      h = track.bottom
    return tracks
}
\end{Verbatim}
\color{black}
\endgroup
\vspace{-0.5em}
\caption{Pseudocode for the grouping and layout algorithm.}
\label{fig:pseudocode}
\vspace{-0.5em}
\end{figure}

\iffalse
\begin{figure}[t]
\begingroup
\scriptsize
\setlength{\baselineskip}{0.5\baselineskip}
\begin{verbatim}
Timeline_Layout(events, W) {
    // Cluster dated Events (DBSCAN)
    d = min(d_base, (2.5 yrs / span) * 100)
    C = Cluster_Dates(dated(events), d, minPts=1)

    // Optimize non-temporal / temporal split
    bestHeight = infinity
    rFinal = 10%
    for each r in [10%, 15%, ..., 50%]:
      h = Draft_Layout(events, W, C, r)
      if h < bestHeight:
        bestHeight = h
        rFinal = r
    Draft_Layout(events, W, C, rFinal)
}

Draft_Layout(events, W, C, r) {
    // Width by event count
    for each s in [C: Past, C1..Ck, Present]:
      s.width = |s| / |all temporal events|
      s.width = s.width * (W * (1 - r))

    // Segment bounds
    x = W * r
    for each s in [C: Past, C1..Ck, Present]:
      s.left = x
      s.right = s.left + s.width
      c.scale = Map(dateRange(c), c.left, c.right)
      x = s.right

    // Track layout (first-fit event stacking)
    for each track in GROUP(events):
      for each e in track:
        lane = 0
        if FITS(lane, e) Place_In_Lane(lane, e)
        else Place_In_Lane(lane++, e)
        compute e.left, e.right
        compute e.top, e.bottom
      track.height = sum(lane heights)
      track.top = previous(track).bottom
      track.bottom = track.top - track.height
}
\end{verbatim}
\endgroup
\vspace{-0.5em}
\caption{Pseudocode for the grouping and layout algorithm.}
\label{fig:pseudocode}
\vspace{-0.5em}
\end{figure}
\fi

The grouping and layout algorithm \texttt{Timeline\_Layout} takes an \added{array of unordered} Events \added{and the window width W in pixels} as input. \added{We} first call \texttt{Cluster\_Temporal\_Events} to form Temporal Clusters. \added{Then many candidate layouts are evaluated through repeated calls to \texttt{Draft\_Layout}. Each candidate uses a different ratio $r$ for the split in widths of the screen space allocated to non-temporal versus temporal data; the split yielding the most compact vertical layout is chosen.} 

\added{The Temporal Clusters created in \texttt{Cluster\_Temporal\_Events} require a global computation, in contrast to} Concern Groups and simple Time Groups (\emph{Unspecified}, \emph{Early}, \emph{Current}) \added{that} are derived directly from Event attributes. \added{We use DBSCAN}~\cite{ester1996dbscan} \added{to capture the punctuated temporal structure of health stories without assuming a fixed number of clusters. In \texttt{Normalize\_Durations} we map the Event durations onto a normalized time axis with the range of 0 to 100. We fix \texttt{eps}, the minimum radius between neighbors, to approximately 2.5 years, permitting more clusters for longer histories than shorter ones. Setting the \texttt{minPts} parameter to 1 allows isolated Events to be singleton clusters. We tuned these parameters with experimentation on the pre-design artifacts collected in Stages 1 through 3.} 
\deleted{Events are mapped onto a normalized time axis from earliest date to most recent and clustered with a minimum of one event per cluster. We adjust DBSCAN's neighborhood radius based on the overall timespan, using a baseline of 30 and a reference gap of approximately 2.5 years, becoming more selective for longer histories and more permissive for shorter ones. We set \texttt{minPts}=1 so isolated dated Events form singleton clusters, and fixed the 2.5-year reference gap used to compute DBSCAN's neighborhood radius after tuning it on pre-design artifacts.}

%and outputs Timescale, Track, and InfoBox coordinates that determine their spatial arrangement. 
\added{In \texttt{Draft\_Layout}, the candidate layouts are created by constructing vertical Segments for the Time Groups and packing Events into horizontal tracks for each Concern Group. The array of Segments starts with the Early group representing the Past and ends with the Current group for the Present, with intermediate items for each of the $k$ Temporal Clusters stored in the previously computed array of clusters \texttt{C}.} 

\added{In \texttt{Position\_Segment}, segments are constructed with widths proportional to the number of dates they contain, to visually emphasize periods of higher temporal density. Within each segment, a linear mapping from the range of dates to the range of available pixels defines the scale factor used to lay out Event InfoBoxes according to the local Timescale.}
\deleted{Segments are constructed with widths proportional to the number of dates they contain, emphasizing periods of higher temporal density. Segments are positioned sequentially from left to right with \textit{s.left} and \textit{s.right} bounds, including \emph{Past} and \emph{Present} segments on either side of Temporal Clusters. Within each segment, a linear mapping from date values to pixel positions defines its local timescale. An Event’s horizontal placement is derived from its StartDate and EndDate, which may span multiple segments. Because each Segment maps only its own date range to its allotted width, inactive spans are condensed while dense periods retain readable space.}

\added{In \texttt{Pack\_Concern\_Tracks}, events are arranged within horizontal Tracks using a first-fit packing strategy. First, Tracks are created with \texttt{GroupByConcern} to associate Concern Groups and their Events with a horizontal section of the display. Tracks laid out from bottom to top. They are divided into horizontal lanes in which InfoBoxes are placed so that they do not overlap. Each Track is initialized with a single lane. To place an InfoBox, the desired horizontal location is computed and the set of lanes is searched from bottom to top to find one where it can be accommodated without overlapping previously placed InfoBoxes. If no existing lane can accommodate the box without overlap, a new lane is created at the top and added to the set, and the box is placed there. The lane height may be increased if the box height requires it. After all the events are placed, the sum of lane heights is the track height, and the next track is placed on top of it.} 
\deleted{Events are arranged within Tracks using a first-fit packing strategy. Each Track is divided into horizontal lanes of non-overlapping InfoBoxes, placed in the first available lane or in a new lane if needed. Track height is the sum of lane heights, and Tracks are stacked vertically by setting the \textit{track.top} of each track to the previous track's \textit{track.bottom} position.}

This approach produces compact visualizations suitable for printing and rapid interpretation.

\begin{figure}[h!]
  \centering
  \includegraphics[width=\columnwidth]{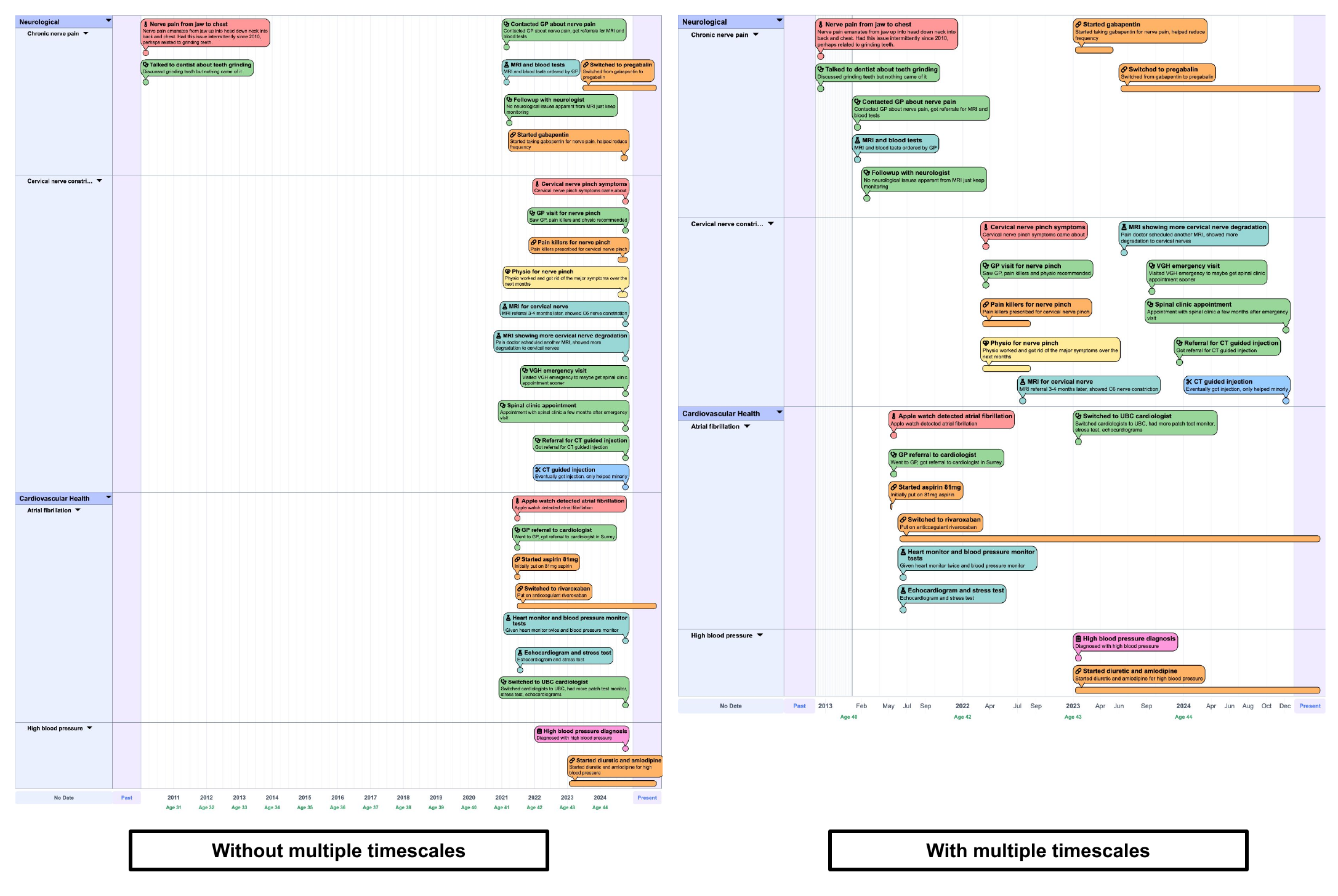}
  \caption{
  Layout comparison for participant E22 from the elicitation sessions, contrasting a uniform single-timescale layout (left) with our multi-timescale approach (right). %The single-timescale version produces a taller visualization with inefficient whitespace usage. In contrast, the multi-timescale layout compacts the structure while preserving temporal context, and more clearly reveals short-duration range markers that are difficult to perceive under a uniform scale.
  }
  \label{fig:layoutcomparison}
\end{figure}

Figure~\ref{fig:layoutcomparison} compares this multi-timescale approach to a single-timescale baseline, which produces taller layouts with inefficient whitespace. Our approach compacts the structure, preserves temporal context, and improves visibility of short-duration ranges.

\subsubsection{Visual Styling Pass}

The elements in the visual encoding are then styled to create the final visual representation, using the alignment and positioning information from the layout computation. 
%the structured data abstraction is mapped to its corresponding visual representation and encoding. Events and groupings are instantiated as InfoBoxes within Tracks, with alignment and positioning derived from the layout, 
Additional marks, including the subtle grid line density variations that convey Timescales, the vertical boundaries of Timescale segments, and the horizontal lines delineating Tracks, are also added in this phase, with their placement determined by calculations from the grouping and layout algorithm section of the pipeline.
\deleted{This level of visual polish ensures that the visualization remains legible as both an interactive interface and a static artifact, supporting quick interpretation without introducing visual clutter.}

\subsubsection{Development}

HealthTale was implemented using \added{TypeScript} \cite{typescript}, \added{Lit} \cite{lit}, and \added{D3} \cite{bostock2011d3}. \added{We incorporated an existing implementation of DBSCAN directly into the codebase.} The system was developed through an iterative prototyping process involving multiple feedback cycles with the research team and our project partner. \deleted{This process ensured that the prototype aligned with both the research goals and the practical constraints of integration within the partner’s ecosystem and clinical workflows.}

We used Claude Code \cite{anthropic2025claude} to assist with the visual styling pass, enabling rapid iteration on visual polish as well as general code cleanup.

\section{Evaluation}

We conducted a multi-stage evaluation of HealthTale combining formative refinement and summative assessment\deleted{, structured around both our system components and design goals}. The evaluation assesses the level of system support for \added{the design goals of} elicitation (DG-E) and communication (DG-C) by targeting four aspects: \textit{data parsing}, \textit{data abstraction}, \textit{grouping and layout algorithm}, and \textit{visual encoding}.

\subsection{Stage 1: Formative System Validation}

We first conducted formative testing using previously collected health story artifacts, namely the unsolicited online narratives from Reddit ($\nOnlineNarratives=20$) and written narratives from our final pre-design study ($\nElicitedStories=22$). The research team provided these artifacts verbatim to HealthTale and reviewed the resulting visualizations \added{to test all four aspects}.

This stage focused on evaluating parser prompting strategies and stress-testing the grouping and layout algorithm across variations in event density, temporal spread, and narrative complexity. We examined whether Events were correctly extracted, whether temporal groupings aligned with expectations, and whether the visualization remained legible under different conditions.

Based on these observations, we refined parser prompts, debugged layout behaviors, and adjusted visual styling details prior to human testing. We saved the visual artifacts and JSON data for use in Stage 4: Clinician Formative Review.

\subsection{Stage 2: Patient Formative Evaluation}

We conducted a user study with patients ($\nPatientEvaluation=34$; age range = 24--65; $M = 35.9$, $Mdn = 31.0$), consisting of 30-minute Zoom or in-person sessions. Participants received a CAD\$20 honorarium and were recruited through snowball sampling and social media posts to capture a range of health complexities. \added{In this $\nPatientEvaluation=34$ study, we used the first 19 sessions for formative refinement in this stage, and the final 15 sessions for summative evaluation in Stage 3.} 

During each session, participants typed their health story into the Elicitation Panel, then considered the output visualization generated by the system and had the opportunity to interact with it via the interface. Participants engaged in a think-aloud process while researchers asked targeted and open-ended questions. We saved their freeform health story, visual artifact and JSON containing the structured data. Fig.~\ref{fig:healthstories} shows a subset of artifacts generated during both formative and summative phases using the final iteration of our visual styling.

\added{For the} first 19 sessions used for formative refinement, \added{we} conducted sessions in an iterative cycle, where participant feedback from interviews informed targeted adjustments to the system, which were then evaluated in subsequent sessions. Iterations were introduced when consistent patterns of feedback emerged across participants, specifically when the same concern was raised repeatedly. These iterations were limited to the visual styling of the system and were intended to improve usability \added{of the visual encoding aspect}. The underlying functionality and utility of the artifact remained unchanged.

\textbf{Iteration 1} ($\nPatientFormative=7$):  
We redesigned the InfoBox styling to resemble dialogue bubbles rather than rectangles.  

\emph{Rationale:} Multiple participants misinterpreted box width as event duration.

\textbf{Iteration 2} ($\nPatientFormative=7$):  
We introduced example health stories prior to elicitation and simplified the editing interface. 

\emph{Rationale:} Multiple participants expressed uncertainty about how much detail to include and found the editing interface overwhelming.

\textbf{Iteration 3} ($\nPatientFormative=5$):  
We adjusted the InfoBox styling to a pill shape and color-matched temporal markers to their boxes. We also refined the elicitation prompt to encourage inclusion of temporal information.

\emph{Rationale:} Multiple participants were misinterpreting the time encoding and confused by the lack of temporal detail.

\subsection{Stage 3: Patient Summative Evaluation}

\added{We used the data from the final 15 of 34 participants in our post-design user study with patients to evaluate HealthTale summatively. To do so, we qualitatively analyzed session transcripts, researcher notes, and visual health story artifacts to identify recurring observations pertaining to all four aspects and the two design goals. The first author conducted the criterion-based analysis, which was iteratively refined through discussion with the last author.}
\deleted{We report findings from the final 15 evaluations.}

\subsubsection{Data Parsing}

Participants found freeform text input natural and easy to use, supporting DG-E by lowering the barrier to constructing health stories. Parsing was generally effective at extracting Events and associated attributes.
\begin{quoting}
\emph{(PT33) “I like the free-form approach, because I appreciated I could just type, and then letting the AI go through and filter out the text itself and make decisions. . . I liked that a
lot more than having to go through [the event panel]. . . ”}
\end{quoting}
However, some confusion remained around distinctions between earlier Events and the Past Segment, partly due to occasional parsing errors.

\subsubsection{Data Abstraction}

Participants reported that the visualization supported recall and organization of past experiences, aligning with DG-E. The inclusion of Life Events was particularly valuable for contextualizing medical histories.
\begin{quoting}
\emph{(PT16) “Especially if I wanted them to know more extensive about my medical history and things I'd gone through and life events, I think that'd be helpful.”}
\end{quoting}

\subsubsection{Grouping and Layout}

The layout accommodated variation in event density and temporal distribution, supporting both DG-E and DG-C by structuring patient narratives while preserving their inherent variability. \added{Participant health stories ranged from a few years to over 50 years, with many spanning multiple decades, demonstrating temporal scalability.} Participants interpreted temporal groupings without requiring precise timestamps, suggesting alignment with how they conceptualize their histories. \added{Patients and clinicians generally found these multiple timescales understandable, even though cross-segment distances are not directly comparable.}

\subsubsection{Visual Encoding}

Participants described the visualization as supporting self-advocacy and improving communication with clinicians, directly supporting DG-C.
\begin{quoting}
\emph{(PT32) “I think sometimes it can be hard when you’re seeing a new doctor, because you’re like, ‘Here’s where I currently am, but here’s all the history and how we got here,’ and that can get kind of brushed over sometimes. So to see it all visualized in the same place, I think would help get them up to speed a little bit faster without feeling like a big burden on myself as the patient.”}
\end{quoting}

They also valued the ability to view their experiences holistically.
\begin{quoting}
\emph{(PT15) “I think this is nice to see as a patient, how you view what’s happening with your body and other health aspects.”}
\end{quoting}

\subsection{Stage 4: Clinician Formative Review}

We conducted three 60-minute semi-structured interviews with clinicians ($\nClinicianEvaluation=3$) from our project partner’s network.

Clinicians reviewed the patient-generated visualizations, regenerated using the most recent styling iteration, with minimal context to simulate real clinical encounters, followed by think-aloud interpretation and guided discussion. \added{Our analysis process was the same as for Stage 3.}

\subsubsection{Data Abstraction}

Clinicians found the abstraction aligned with their needs and emphasized the importance of Life Events, supporting DG-C by enabling a more complete understanding of patient context.
\begin{quoting}
\emph{(CT2, on if the inclusion of Life Events is important) “Oh, absolutely. I see those as being the precursors that a lot of clinicians aren’t looking for.”}
\end{quoting}
\begin{quoting}
\emph{(CT2) “One of the things that often gets missed by clinicians is when they’re not seeing the whole history.”}
\end{quoting}

\subsubsection{Grouping and Layout}

Clinicians reported that visualizations were easy to scan and supported rapid identification of \textit{information scents}~\cite{pirolli2000infoscent}, aligning with DG-C.
\begin{quoting}
\emph{(CT1) “In a typical summary, it’s very hard to piece together what happened when. Timeline is always a really hard thing to piece together, so having a visual depiction like this is really helpful. I feel like that’s always been a limitation on a lot of EMRs, where everything is documented in just, like, blocks of text with timestamps. ”}
\end{quoting}

\subsubsection{Visual Encoding}

Clinicians noted that the visualization supported a broader understanding of the patient beyond clinical records, directly supporting DG-C.
\begin{quoting}
\emph{(CT1) “It gives you a view not only of their medical history, but also of them as a person.”}
\end{quoting}

They also described its usefulness for assessing the level of patient engagement with care.
\begin{quoting}
\emph{(CT1) “This kind of just means that they're on top of their own health.”}
\end{quoting}

Based on clinician feedback, we made very minor styling adjustments to improve the clarity of hierarchical grouping by strengthening visual boundaries between levels of Concern Groups.

\section{Discussion}

\deleted{Our findings highlight the importance of supporting both elicitation (DG-E) and communication (DG-C) in patient-generated health stories.} Across evaluations, narrative-first input enabled patients to construct representations that reflect their own understanding of their health, while structured visualizations made these representations interpretable in clinical contexts. %Together, these results demonstrate that patient-generated narratives can be effectively structured to support meaningful clinical use. 

These promising outcomes have motivated our project partner to explore integrating the system into their environment, which would enable more extensive evaluation within real clinical workflows.

\subsection{Elicitation, Autonomy, and Holistic Understanding}

\deleted{HealthTale's ability to support elicitation (DG-E) is a crucial first step: patients must first construct a coherent account of their health before it can be effectively communicated.} Our findings highlight that enabling patients to externalize their experiences in a structured yet flexible way helps support both reflection and communication.

Participants consistently described HealthTale as a tool for self-advocacy. Many noted that it is difficult to convey the full extent of their experiences during clinical encounters, particularly when histories are long, complex, or fragmented across multiple providers. By constructing a visual representation of their health story, patients externalized patterns, duration, and severity in ways that are difficult to communicate verbally. This externalization shifts the burden of recall and explanation from real-time conversation to pre-constructed representation, enabling patients to arrive at encounters better prepared to communicate their experiences.

Participants also emphasized the importance of maintaining ownership over their health narrative. Unlike EMRs, where patients are often uncertain what information clinicians can see or how they interpret it, HealthTale allows patients to curate and present a complete account of their experiences. It supports a shift from passive reporting to active authorship, where patients can ensure that they include and communicate relevant context. Importantly, this authorship is not only about completeness, but about framing. Patients are able to decide what is important to tell and how it is told.

Surprisingly, while HealthTale supports editing and refinement, participants engaged in less curation than we anticipated. Past work from ten years ago on curating freeform text into visual timelines emphasized the need of iterative user correction of imperfect natural language processing (NLP) results~\cite{fulda2016timeline}. However, in our current studies, we found that users rarely made substantial changes to the automatically generated visualization, suggesting that the initial structuring was generally sufficient for their needs. Across datasets, we observed relatively few parsing errors, 17/320 events from Reddit, 6/345 from the elicitation sessions, and 18/425 from the patient testing, with a total of 41/1090 misparced events, indicating generally robust performance. Although these results are generally a positive sign that NLP has passed a crucial usability threshold, this limited engagement with curation introduces a challenge: when errors occur, they often go uncorrected. For example, we observed cases where Events were incorrectly placed into the Past segment despite containing temporal signals that should place them in a Temporal Cluster on the timeline, yet participants did not notice or adjust these misplacements.

An outcome that we found especially compelling is that from these visualizations, clinicians were able to infer clinically relevant aspects of the patient's personality that were richer than straightforward medical information such as diagnoses and timelines. Examples mentioned by the clinicians were assessing whether a patient was likely to adhere to a stringent treatment regimen, whether patients were symptom- or treatment-focused, and how well patients understood their own medical condition. These inferences were not only driven by the presence of contextual information, but by what patients chose to include and how they structured it. Patterns such as frequent provider visits, detailed tracking, or emphasis on certain types of Events signaled how engaged, attentive, or systematic patients were in managing their health. Participants’ selections and organization of Events revealed personal priorities, implicit reasoning, and approaches to care that are not well captured in EMRs. This information is not extraneous; it provides critical insight into how patients engage with and manage their health, highlighting the need for tools that preserve and make these signals interpretable in clinical contexts.

Our findings also surface a tension between patient-authored narratives and structured clinical data, suggesting a potential role for integration with existing systems. While HealthTale prioritizes patient-authored narratives, selectively incorporating structured information, such as known diagnoses or timestamps from EMRs, could help anchor Events temporally and reduce misclassification. However, we must integrate such data carefully. Over-integration risks shifting the system away from a patient-centered narrative toward a clinician-centered record. Instead, EMR data may be most valuable as a scaffolding mechanism during elicitation, supporting accuracy while preserving patient control over interpretation and inclusion.

\subsection{Communicating, Shared Understanding, and Efficiency}

\deleted{HealthTale's ability to support communication (DG-C) allows clinicians to more rapidly interpret and act on patient information during time-constrained encounters.} %Our findings highlight that providing structured, interpretable representations enables efficient communication and alignment of understanding between patients and clinicians.

From a clinical perspective, the visualization supports rapid orientation and shared understanding. Clinicians described using the timeline to get a high-level understanding at the beginning of an encounter, enabling them to quickly identify relevant Events and areas for further inquiry. This ability to establish a “lay of the land” is particularly valuable in time-constrained settings, where clinicians must quickly form a mental model of a patient’s history.

The timeline structure also provides a shared reference point for communication. Clinicians noted that they could point to specific Events or time periods and ask patients to elaborate, supporting more targeted and efficient conversations. Rather than reconstructing a patient’s history through questioning alone, the visualization enables a more collaborative interaction, where both patient and clinician can reference the same external representation. This approach reduces ambiguity and helps align mental models between the two parties.

Although HealthTale provides filtering and collapsible track functionality that could be used to reduce visual complexity, patients consistently preferred to keep all information visible in the static artifact they would use during clinical encounters. Participants stated that even details they perceived as less relevant might be important to the clinician, and thus favored presenting a complete view and selectively discussing key points rather than pre-filtering the representation. These decisions suggest that completeness, rather than minimalism, is critical in supporting effective communication.

Importantly, the visualization supports a more holistic understanding of the patient. By including Life Events alongside clinical information, clinicians were able to interpret health trajectories in context, such as identifying stressors or environmental factors that may influence outcomes.

Again, the potential of tighter integration between HealthTale and existing clinical systems presents both an opportunity and a challenge. EMR data could provide temporal anchors or validation signals that improve the reliability of the visualization, helping clinicians trust the representation as a summary of the patient’s history. At the same time, clinicians expressed that the value of HealthTale lies in its ability to surface information not typically captured in EMRs. As such, the goal is not to merge these systems into a single representation, but to allow them to complement one another, where EMRs provide structured, clinically validated data, and HealthTale provides contextualized, patient-authored narratives.

\deleted{Together, these findings suggest that structuring patient-generated narratives can help bridge the gap between patient and clinician mental models. By supporting both expressive input and interpretable output, HealthTale enables patients to communicate their experiences more effectively while allowing clinicians to engage with those experiences in a structured and efficient manner.}

\subsection{Implications of LLM-Supported Data Abstractions}

Another notable insight that arose from our work is in how LLMs enable a shift in how visualization systems acquire and structure data. Rather than requiring users to conform to predefined schemas, systems can now accept freeform input and extract structured representations from how people naturally describe their experiences. This fluidity can lower the barrier to data entry and makes it possible to capture forms of data that were previously difficult to formalize, particularly those that are narrative, contextual, or loosely specified.

This shift places new emphasis on the design of data abstractions. As users provide input in their own words, the resulting data is inherently ambiguous, incomplete, and unevenly specified. Our data abstraction reflects this reality by explicitly accommodating imprecision and incorporating non-seemingly relevant context such as Life Events. Rather than enforcing strict standardization, the abstraction preserves how individuals express and structure their experiences, while still enabling organization and interpretation.

More broadly, this shift suggests a potential transition away from rigid, precision-oriented data abstractions toward more flexible representations that can accommodate the variability of human expression. In this paradigm, the role of user-generated narrative visualization systems is not only to display structured data, but to mediate between ambiguous, user-generated input and interpretable representations. While this paradigm introduces challenges, such as less transparency between the input and output, it also creates opportunities to design systems that surface uncertainty, support refinement, and better align with how people naturally think and communicate.

Although the explorations we report here occur solely in the context of health stories, these implications do extend to other domains where user-generated, narrative data is used. As LLM-supported pipelines become more common, the need for abstractions that balance flexibility with interpretability will become increasingly important.

\subsection{Limitations and Future Work}

This work has several limitations. First, although we conducted many rounds of evaluation, the system has not yet been evaluated in a full clinical setting. Participants created their health stories within a study context, and several noted that they would likely include more detailed information when preparing for an actual appointment. As a result, our findings may underestimate the richness and density of health stories in practice.

Also, while HealthTale is designed to support both the elicitation and communication of health stories, our validation to date primarily focuses on the elicitation process. Our findings show that patients can effectively construct structured representations and benefit from recall, organization, and self-advocacy, with initial evidence that these representations support communication. Future work should more directly evaluate how these representations are used in real clinician–patient interactions and how they influence communication, decision-making, and shared understanding.

Moreover, \added{our clinician sample was small and} were recruited from our project partner’s clinical team. While this approach allowed us easy access to domain-relevant feedback, these clinicians may be more technologically inclined and familiar with the broader project context than the general population of healthcare providers.

Additionally, future work should explore how patient-generated narratives can be integrated with clinical systems in a way that supports both accuracy and expressiveness. While our work focuses on initial clinical encounters, EMRs play a critical role in long-term record maintenance. An important direction is determining how structured clinical data can be selectively incorporated to support temporal grounding and continuity while preserving patient authorship and contextual richness, alongside developing grouping and layout strategies that scale to larger, continuously curated health stories.

\section{Conclusion}

We present HealthTale, a patient-centric health story visualization system that supports elicitation and communication of patient-generated narratives. We introduce a data abstraction that models health stories as Events organized across multiple dimensions, capturing both clinical information and ambiguous lived experience, grounded in a multi-stage qualitative investigation. 

Through evaluation with patients and clinicians, we demonstrate that structuring health stories into visual representations can support patient recall, organization, and self-advocacy, while enabling clinicians to more quickly interpret patient histories and develop a shared understanding. These findings highlight the value of patient-generated narrative data as a complementary source of information within clinical contexts.

Overall, this work suggests that tools that support patients in creating and communicating their health stories offer a promising approach to bridging the gaps between patient and clinician perspectives, enabling more efficient, contextualized, and patient-centered care.

\section*{Acknowledgements}

We are grateful to Thrive Health, and particularly Diane Tam, for their continued collaboration and support of the project. We thank Nureya Khimani, William Hu, and Connie Jorsvik for contributing their clinical expertise to the evaluation of the tool. We also acknowledge the use of Claude Sonnet 4.6 to assist with coding aspects of the system, particularly the implementation and refinement of its visual polish and interactive features. We thank Steve Kasica, Matt Oddo, and Mara Solen of the UBC InfoVis Group for their feedback. This work was funded by Thrive and Mitacs through the Mitacs Accelerate program.

\section{Supplementary Materials}

Supplementary materials can be found at \url{https://osf.io/t75qn}. We provide the scripts and protocols for all human-subjects studies and the demographics of the participants. We provide the health story artifacts of the early patient roleplay interviews,  written elicitation sessions and the collected Reddit posts, the results of visualizing these written stories in the final version of HealthTale, and all visualization results from the evaluation sessions where patients directly used HealthTale. We also provide the JSON datasets with parsed Events corresponding to all visualization results. 

We also provide the intermediate analysis artifacts of structured decompositions of the Reddit posts and elicitation session results, and our thematic analysis of the patient and clinician interviews. We provide the LLM parser prompts used in HealthTale, a more detailed version of the grouping and layout algorithm pseudocode, and a complete list of all cases where the parser incorrectly categorized events in the summative evaluation sessions. We provide the full codebase, released under the BSD open source license. 

Finally, we have a PDF and video walkthrough to show the look and feel of a HealthTale usage session. The video walkthrough can also be viewed at \url{https://youtu.be/kwQqRZV3n_Q}.

\bibliographystyle{main/abbrv-doi-hyperref}

\bibliography{main/main}

\end{document}